\documentclass[12pt,preprint]{aastex}

\usepackage{natbib}
\shorttitle{What happened to the other Mohicans?}
\shortauthors{Garaud, P.}

\begin{document}

\title{What happened to the other Mohicans? \\ 
Realistic models of metallicity dilution by fingering convection and observational implications.}


\author{P. Garaud$^{1,2}$}
\affil{$^1$ Department of Applied Mathematics and Statistics, Baskin School of Engineering, UC Santa Cruz, 1156 High Street, Santa Cruz CA 95064 \\ 
$^2$ On sabbatical leave at: Institute for Astronomy, 34 `Ohi`a Ku St., Pukalani, HI 96768-8288}


\begin{abstract}
When a planet falls onto the surface of its host star, the added high-metallicity material does not 
remain in the surface layers, as often assumed, but is diluted into the interior
through fingering (thermohaline) convection. Until now, however, the timescale 
over which this process happens remained very poorly constrained. 
Using recently-measured turbulent mixing rates for fingering convection, 
I provide reliable numerical and semi-analytical estimates for the rate at which 
the added heavy elements drain into the interior. I find that the {\it relative} metallicity 
enhancement post-infall drops by a factor of ten over a timescale 
which depends only on the structure of the host star, and decreases very rapidly 
with increasing stellar mass (from about 1Gyr for a 1.3$M_\odot $ star to 10Myr for a 1.5$M_\odot$ star). 
This result offers an elegant explanation to the lack of observed trend between metallicity and 
convection zone mass in planet-bearing stars. More crucially, it strongly suggests that the statistically-higher metallicity of planet-bearing stars must be of primordial origin. Finally, the fingering region is found to extend 
deeply into the star, a result which would provide a simple theoretical explanation to the 
measurements of higher lithium depletion rates in planet-bearing stars. 
\end{abstract}




\section{Introduction}

Transit observations have recently revealed the existence of a number
of very short period exoplanets \citep{Torresal2008}. Theoretical investigations 
\citep{JBG2009,LWC2009} show that most of them are tidally unstable and doomed to fall onto the central star. 
These recent findings confirm the notion that, through disk migration followed by tidal interactions, the vast majority of planets formed may not actually survive \citep{LinPapaloizou1986}. Back in the 1990s, this idea led Doug Lin to coin the colorful phrase ``the last of the Mohicans'', to describe the few which -- by one process or another -- escape the fate of their siblings. 

The directly-related question of whether the statistically-higher metallicity of planet-bearing stars \citep{FischerValenti2005} results from planetary infall or is of
primordial origin remains unanswered to date. It is of crucial importance, however, since it bears on the debate between the core-accretion and gravitational-instability scenarios of planet formation. The various ideas proposed to find evidence in favor of the planetary infall theory are typically based on the premise that the added material is mixed with the outer convection zone only. 
The most commonly-discussed one relies on the very rapid decrease of the mass of the outer convection zone with stellar mass between $1M_\odot$ and $2M_\odot$. The relative metallicity enhancement caused by the infall of similar-sized planets should therefore be much larger for higher-mass stars than for solar-mass stars \citep{LaughlinAdams1997}.  This effect, however, has never been detected \citep{SIM2001,FischerValenti2005}, 
a fact which may seem rather odd now that we know that infall events must regularly happen. Hence the question:  what happened to the ``other'' Mohicans? 

\citet{vauclair2004} was the first to question the original assumption that the added heavy-elements remain in the convection zone. She noted that, in the presence of adverse $\mu$-gradients, a stably-stratified region could be destabilized and become turbulent through a well-known double-diffusive fingering instability, often called, by analogy with the oceanic context, ``thermohaline convection'' \citep{stern1960}. She argued that the induced mixing could slowly dilute the excess metallicity into the underlying radiative zone.

Whether a system is stable or unstable to the fingering instability depends on the so-called density ratio $R$, defined in astrophysics \citep{ulrich1972} as:
\begin{equation}
R =  \frac{\nabla - \nabla_{\rm ad} }{ \nabla_\mu}  \mbox{  , } 
\end{equation} 
where, as usual,  $\nabla - \nabla_{\rm ad} =\left(\frac{\partial \ln T}{\partial \ln p}\right) -  \left(\frac{\partial \ln T}{\partial \ln p}\right)_{\rm ad} $ and $\nabla_\mu = \left(\frac{\partial \ln \mu}{\partial \ln p}\right) $, $p$ and $T$ are pressure and temperature, and $\mu$ is the mean molecular weight. 
The fingering instability takes place when 
\begin{equation}
1 < R  < \tau^{-1} \mbox{   where   } \tau = \frac{\kappa_\mu}{\kappa_T} \mbox{  , } 
\end{equation}
$\kappa_T$ is the thermal diffusivity and $\kappa_\mu$ is the compositional diffusivity. A region with $R < 1$ is Ledoux-unstable to direct overturning convection. The fingering instability occurs in Ledoux-stable systems, and is stabilized only when $R > 1/\tau$. Since $1/\tau$ is typically of the order of $10^6$ to $10^7$ in stars, \citep{ulrich1972,schmitt1983}, even a very modest adverse compositional gradient can trigger the instability. 
\citet{vauclair2004} tentatively estimated that, through the added mixing, the decay timescale of the surface metallicity enhancement post-infall should be about 1,000 yrs, which would rule out the possibility of detection. At the time, however, fingering convection in the astrophysical context was much more poorly understood, and little was known about the strength of the turbulence actually induced. 

This, however, is no longer the case: using 3D numerical simulations, \citet{Traxleral2010} recently quantified turbulent transport by fingering convection in astrophysics. 
Their mixing laws are summarized in \S\ref{sec:mixing}.  I apply them, in \S\ref{sec:example}, to study the evolution of the $\mu-$profile of a 1.4$M_\odot$ star, polluted by a high-metallicity impactor. The results confirm most aspects of Vauclair's original idea but are now quantitatively reliable. In \S\ref{sec:estim} I propose very simple estimates of the process, which can be applied to a wide range of stars and impactors without the need for heavy-duty numerical integration. I conclude in \S\ref{sec:ccl} by discussing the fundamental implications of these findings.

\section{Turbulent mixing by fingering convection}
\label{sec:mixing}

\citet{Traxleral2010}  found that compositional transport induced by fingering convection is accurately modeled through the following diffusion coefficient:
\begin{equation}
D^{\rm finger}_\mu = \sqrt{\nu \kappa_\mu } g(r) \mbox{  , } 
\label{eq:dfinger}
\end{equation}
where $\nu$ is the viscosity, 
\begin{equation}
r = \frac{R - 1}{1/\tau - 1} \mbox{  , } 
\label{eq:r}
\end{equation}
is a reduced density ratio\footnote{  $R \in [1,1/\tau]$ corresponds to $r \in [0,1]$. The lower limit is now $r=0$ (below which the system is Ledoux-unstable) and upper limit is $r = 1$ (above which the system is fingering-stable).},
and $g(r)$ is a universal function fitted from the numerical data:
\begin{equation}
g(r) = 101 e^{-3.6r} (1-r)^{1.1} \mbox{   if   } r \in (0,1) \mbox{   and   }  g(r) = 0 \mbox{ otherwise. } 
\end{equation}
It is crucial to note that (\ref{eq:dfinger}) has been obtained through experimental measurements, with no additional modeling, and contains no free parameters: it is a universal law for mixing by fingering convection.
\citet{Traxleral2010} also found that the induced turbulent heat transport is negligible compared with the molecular and radiative contributions \citep{Denissenkov2010}. Hence it does not affect the background entropy profile of a star.

In what follows, I adopt the following algorithm to calculate the total compositional diffusivity $D_\mu(x,t)$: 
\begin{itemize}
\item For a given $\mu(x)$ profile (where $x$ is the distance to the stellar center normalized by the stellar radius $R_\star$), calculate $r$.
\item If $r > 1$, set $r$ to be 1 while if $r < 0$ set $r$ to be 0. 
\item Set $D_\mu(x,t) = D_\mu^{\rm conv}(x,t) + D_\mu^{\rm finger}(x,t) + \kappa_\mu(x)$. 
\end{itemize}
In this last expression $D_\mu^{\rm conv}(x,t)$ is the fully convective diffusivity, chosen to be: 
\begin{equation}
D_\mu^{\rm conv}(x) = \frac{D^{\rm conv}}{2} \left[ 1 + \tanh\left( \frac{x - x_{\rm L}(t)}{\Delta} \right) \right] \mbox{   , }
\label{eq:dconv}
\end{equation}
where $x_{\rm L}$ marks the transition between the Ledoux-unstable and the Ledoux-stable regions (i.e. below which $r$ is strictly positive), $\Delta$ is the assumed thickness of the overshoot region (here $\Delta/R_\star = 0.001$), and $D^{\rm conv}$ is a very large convective value derived from mixing-length theory. 

\section{Example of evolution of a polluted system}
\label{sec:example}

\subsection{The Model} 

I now apply this new mixing law to model the evolution of the $\mu$-profile in a 1.4$M_\odot$ star recently impacted by a $\sim 1 M_J$ planet. The background stellar model used was provided by P. Bodenheimer, see \citet{GaraudBodenheimer2010} for detail. 
The background mean molecular weight\footnote{I assume {\it for simplicity} that, in all objects, the Helium to Hydrogen abundance ratio is constant with $Y/X = 0.39$, and that the material is fully ionized, so that $\mu^{-1} \simeq (1.65 - 1.15Z)$.} assumed is $\mu_\star \simeq 0.61$. 
Since the total mass of the outer convection zone of this star is significantly smaller than 1$M_J$, the metallicity of the polluted envelope is determined by that of the infalling planet. Here I choose a high-metallicity case with an initial molecular weight $\mu_{\rm 0} \simeq 0.64$ in the convection zone. 

In what follows, I compare the results of two related simulations, one in which fingering convection 
is ignored (by setting $D_\mu^{\rm finger}$ to 0), and one in which it is taken into account. In both cases, all parameters, initial conditions and backgrounds are otherwise identical. The value of $D^{\rm conv}$ (see (\ref{eq:dconv})) is taken to be equal to $10^{8}$cm$^2$/s which is a few orders of magnitude smaller than commonly-used convective diffusivities. This choice merely facilitates the numerical computation, as the timesteps would otherwise have to be prohibitively small. Indeed, it can be shown that the only effect of selecting a smaller convective diffusivity is to increase the timescale for the transient dynamical readjustment phase (see \S\ref{sec:estim}). 

The initial $\mu$-profile is:
\begin{equation}
\mu(x) = \mu_\star + \frac{\mu_0-\mu_\star}{2} \left[ 1 + \tanh\left(\frac{x-x_{\rm cz}}{\Delta}\right)\right]  \mbox{   ,  }
\end{equation}
where $x_{\rm cz}$ marks the base of the convection zone prior to infall. It then evolves according to:
\begin{equation}
\frac{\partial \mu}{\partial t} = \frac{1}{ \rho x^2 R_\star^2} \frac{\partial}{\partial x} \left( \rho x^2 D_\mu(x) \frac{\partial \mu}{\partial x} \right)   \mbox{  ,}
\end{equation}
where $\rho$ is the background density, $D_\mu(x)$ is given by (\ref{eq:dfinger}), where $r$ is given by (\ref{eq:r}) and $R$ is rewritten more explicitly as a function of $\mu(x)$ via:
\begin{equation}
R(x)  = \frac{ \mu(x) N^2(x) R_\star }{ g(x) \frac{\partial \mu}{\partial x}} \mbox{  , } 
\label{eq:bigRdef}
\end{equation}
where $g(x)$ is the local gravity and $N^2 = (g/H_p) (\nabla_{\rm ad} - \nabla)$ with $H_p$ the pressure scaleheight.  The functions $\rho(x)$, $g(x)$ and $N(x)$, as well as $\nu(x)$, $\kappa(x)$ and $\kappa_\mu(x)$ are extracted from the stellar model. 

Both simulations are evolved from the time of infall $t=0$ up to $t = 700$Myr.  

\subsection{Results}

The evolution of the $\mu$-profile goes through two phases, as originally discussed by \citet{vauclair2004}: a first rapid dynamical readjustment phase, and a second much slower fingering phase. These are now described in more detail. 

\begin{figure}
\epsscale{0.5}
\plotone{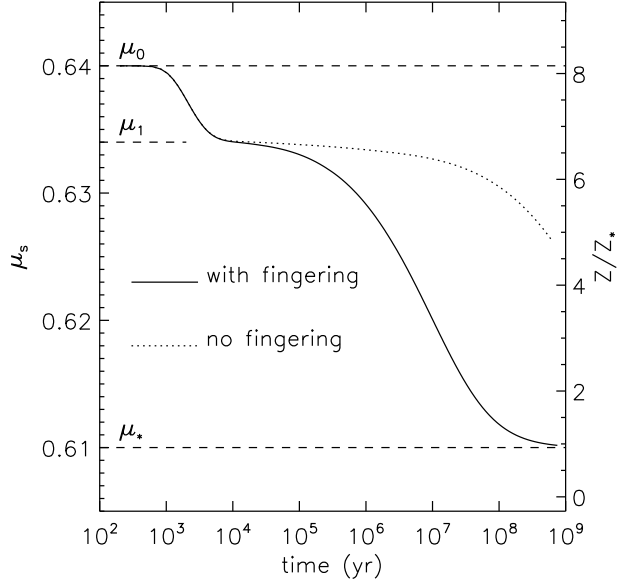}
\caption{Evolution of the surface mean molecular weight $\mu_s(t)$ of both simulations described in \S\ref{sec:example}. }
\label{fig:musurf}
\end{figure}

In the rapid adjustment phase (here $ t< 10,000$yr), the depth of the outer convection zone first increases as the system becomes Ledoux-unstable. 
Convective mixing with the underlying material dilutes the added metals, and the surface mean molecular weight decreases. As mixing proceeds, the compositional gradient at the base of the convection zone softens, and the convective zone retreats back to its original position, leaving behind a region of enhanced metallicity. 
After the end of this dynamical phase, the mean molecular weight of the convection zone is again more or less constant, but with a value $\mu_1$ smaller than the original $\mu_0$. The region near the base of the convection zone is now marginally stable to the Ledoux-criterion. These points are illustrated in Figure \ref{fig:musurf}. Note how up to now the difference between the models with and without fingering is practically indistinguishable. 

\begin{figure}
\epsscale{0.5}
\plotone{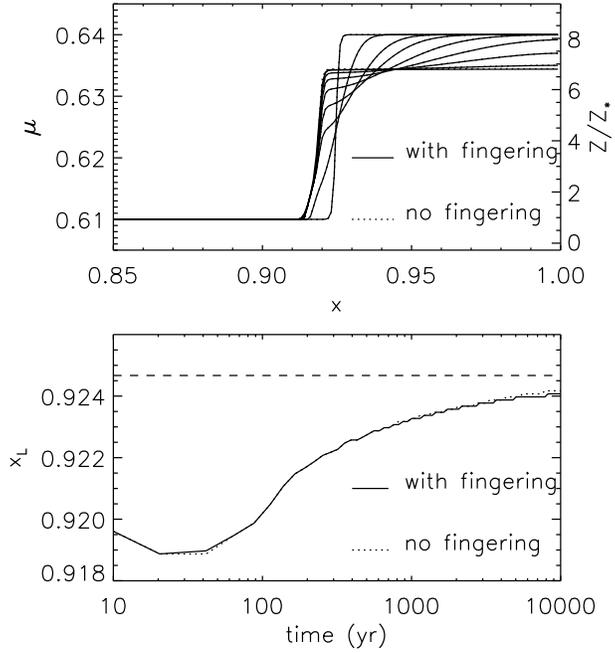}
\caption{Top panel: Evolution of $\mu(x)$ (equivalently $Z/Z_\star$) in the early dynamical adjustment phase of the outer convection zone. 
Shown are representative snapshots at (from top to bottom line in the convection zone) $t = 0$, $t=40$yr, $t=160$yr, $t=320$yr, $t=640$yr, $1,200$, $t=2,400$yr, $t = 4,800$yr and $10,000$yr.  The results of the simulations with fingering are indistinguishable from the one without. Bottom panel: early evolution of the position of the base of the convectively unstable region, $x_L$. Its minimum is called $x_{\rm mix}$ in \S\ref{sec:estimdyn}. The dashed line marks $x_{\rm cz}$ of the unpolluted star. }
\label{fig:earlymu}
\end{figure}
 
\begin{figure}
\epsscale{0.5}
\plotone{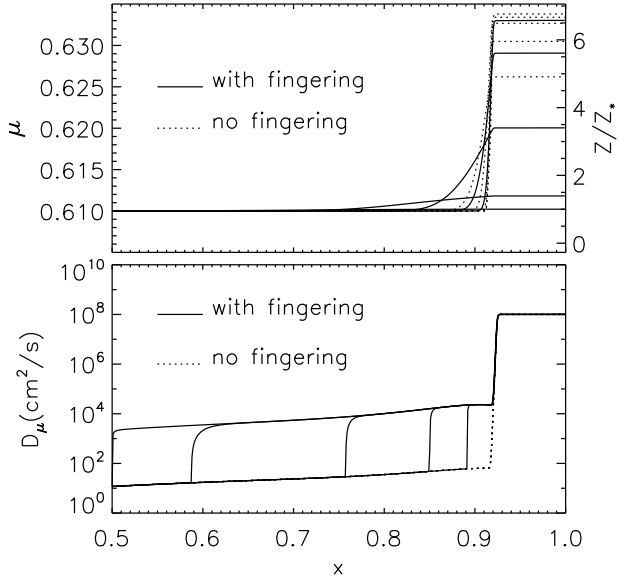}
\caption{ Evolution of $\mu(x)$ and $D_\mu(x)$ for the simulations described in \S\ref{sec:example}. Shown are times (from top to bottom line in the convection zone, top panel and from right to left in the fingering region, bottom panel) $t = 10^5$yr, $t=10^6$yr, $t=10^7$yr, $t= 10^8$yr and finally $t = 7 \times  10^8 $yr. At $t = 7 \times  10^8 $yr, the fingering region spans the entire computational domain. }
\label{fig:latemu}
\end{figure}

Mixing by fingering convection begins to affect the system after about $10^5$ years. The flux of high-$\mu$ material destabilizes the top of the stably-stratified region. The fingering region, which can be seen in Figure \ref{fig:latemu} as the extended interval where $D_\mu \sim 10^4$cm$^2$/s, slowly but steadily deepens with time\footnote{Its progression would only be stopped where the background $\mu$ increases again in the nuclear burning region, not included in the computational domain $x\in[0.5,1]$.}.
At these stellar parameters, the turbulent compositional transport is about a hundred times more efficient than molecular transport. By $t \sim 100$Myr, the surface metallicity has nearly returned to its original value in the fingering case, but decays much more slowly in the absence of fingering.

\section{Estimated timescales}
\label{sec:estim}

The simulation presented above serves as a detailed illustration of the basic two-stage dilution process. However, 
one can also roughly approximate these two stages according to the following estimates, which can be easily applied to a wide class of stars without the need for numerical integration. 

\subsection{Dynamical phase}
\label{sec:estimdyn}

\begin{figure}
\plotone{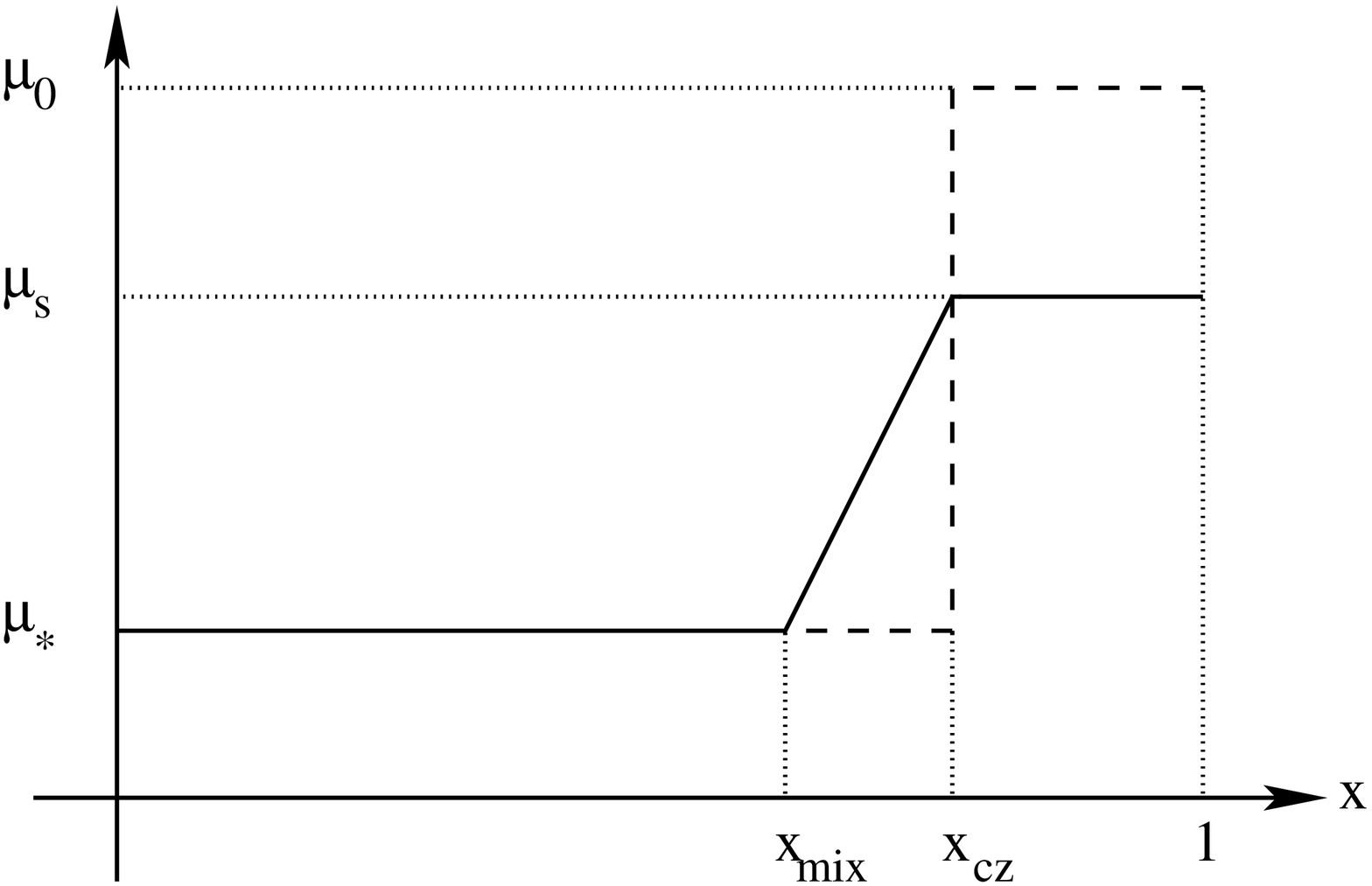}
\epsscale{0.5}
\caption{Schematic of the simplified model. The dashed line shows the initial condition, and the solid line the current profile. The convection zone is homogeneous, with mean molecular weight $\mu_s$. The radiative region adjacent to the convection zone is partially mixed, with a uniform gradient between $x_{\rm cz}$ and $x_{\rm mix}$. In \S\ref{sec:estimdyn} $\mu_s = \mu_1$; in \S\ref{sec:estimfing} $\mu_s$ and $x_{\rm mix}$ evolve with time. }
\label{fig:dilution}
\end{figure}

The end-product of the dynamical dilution phase can be modeled in the manner depicted in Figure \ref{fig:dilution}, which is inspired from the results discussed in \S\ref{sec:example}. The convection zone has retreated back to its original position $x_{\rm cz}$, and has a mean molecular weight $\mu_s = \mu_1$. The compositional gradient in the region below has been smoothed out down to the lowest point of excursion of $x_{\rm L}$, which I call $x_{\rm mix}$. I assume that $d\mu/dx$ is constant between $x_{\rm mix}$ and $x_{\rm cz}$, and therefore equal to $(\mu_1 - \mu_\star)/(x_{\rm cz}- x_{\rm mix})$. 

The convection zone dilution process conserves the total mass of added heavy elements so that,  approximately, 
\begin{equation}
(\mu_0 - \mu_\star) M_{\rm cz}  = (\mu_1 -\mu_\star) \frac{M_{\rm mix} + M_{\rm cz}}{2} \mbox{   ,}
\end{equation}
where $M_{\rm cz}$ is original mass of the outer convection zone, and $M_{\rm mix}$ is the mass of the spherical shell above $x_{\rm mix}$. This yields $\mu_1$ as:
\begin{equation}
\mu_1 = \mu_\star + \frac{2M_{\rm cz}}{M_{\rm mix} + M_{\rm cz}} (\mu_0 - \mu_\star) \mbox{   .}
\label{eq:mu1}
\end{equation}

The convection zone starts retreating when the system becomes Ledoux-stable, hence the marginal stability condition $R=1$ uniquely determines $x_{\rm mix}$. 
At $x_{\rm mix}$, $\mu = \mu_\star$, and $d\mu/dx$ is roughly equal to a half the value it has in the region between $x_{\rm mix}$ and $x_{\rm cz}$ (since it is zero below $x_{\rm mix}$). Using (\ref{eq:bigRdef}) and (\ref{eq:mu1}), the condition $R=1$ becomes:
\begin{equation}
\frac{\mu(x_{\rm mix} ) N^2 (x_{\rm mix}) R_\star}{ g(x_{\rm mix})  \left. \frac{d\mu}{dx}\right|_{x_{\rm mix}} }= \frac{2 \mu_\star N^2(x_{\rm mix}) R_\star}{ g(x_{\rm mix}) \frac{2M_{\rm cz}}{M_{\rm mix} + M_{\rm cz}}  \frac{\mu_0 - \mu_\star}{x_{\rm cz}- x_{\rm mix}} } = 1  \mbox{  . }
\label{eq:xmix}
\end{equation}  
This equation can be solved for $x_{\rm mix}$, which then yields $\mu_1$.  For the 1.4$M_\odot$ 
stellar model used here, the solution of (\ref{eq:xmix}) is $x_{\rm mix} = 0.9165$, so $\mu_1 = 0.636$. This turns out to be a satisfactorily accurate estimate of the numerical results of \S\ref{sec:example} (where $x_{\rm mix} = 0.919$ and $\mu_1 = 0.634$, see Figure \ref{fig:musurf}). 

Another test of this method  was made integrating the evolution of the $\mu-$profile of a 1.3$M_\odot$ star with $\mu_\star = 0.61$, subject to a smaller initial enhancement ($\mu_0 = 0.615$) since its convection zone mass is now comparable with a Jupiter mass. The solution of equation (\ref{eq:xmix}) yields $x_{\rm mix} = 0.8695$ and $\mu_1 = 0.6147$, which is again remarkably close to the numerical solution ($x_{\rm mix} = 0.8705$ and $\mu_1 = 0.6144$). 

\subsection{Fingering phase}
\label{sec:estimfing}

Given that even a very small adverse compositional gradient can destabilize the stratified background, the fingering region propagates rapidly into the interior. Moreover, within the bulk of this region, the turbulent compositional diffusivity remains close to its maximum value $D^{\rm finger}_{\mu,{\rm max}} = 101 \sqrt{\nu \kappa_\mu} $, as seen in Figure \ref{fig:latemu}. Based on these observations, I model the dilution of the convection zone heavy-elements into the radiative interior in the following way. 

As in \S\ref{sec:estimdyn}, I consider a smoothly-mixed region, lying between $x_{\rm mix}$ and $x_{\rm cz}$, where $x_{\rm mix}$ is now assumed to deepen with time through a slow diffusion process. The evolution of $x_{\rm mix}(t)$ is derived by setting the diffusion timescale (via fingering) across the interval $[x_{\rm mix},x_{\rm cz}]$ equal to the time elapsed since infall, $t$:
\begin{equation}
x_{\rm mix}(t) = x_{\rm cz}  - \sqrt{ \frac{\alpha D^{\rm finger}_{\mu,{\rm max}}  t }{R_\star^2} } \mbox{   ,} 
\end{equation}
where $\alpha$ (which can be thought of as a spatial eigenvalue of this diffusion problem) is a number of order one which will be determined more precisely later, and $ D^{\rm finger}_{\mu,{\rm max}} $ is estimated from a typical value near $x_{\rm cz}$. As in \S\ref{sec:estimdyn}, by mass conservation:
\begin{equation}
\mu_s(t) = \mu_\star + \frac{2M_{\rm cz}}{M_{\rm cz} + M_{\rm mix}(t)} (\mu_1- \mu_\star) \mbox{   ,} 
\label{eq:mus}
\end{equation}
where $\mu_s(t)$ is the evolving surface mean molecular weight, $\mu_1$ is its value at the end of the dynamical readjustment phase (estimated via the solution of  (\ref{eq:mu1}) and (\ref{eq:xmix}), see \S\ref{sec:estimdyn}), and $M_{\rm mix}(t)$ is the mass of the spherical region above $x_{\rm mix}(t)$. 

\begin{figure}
\epsscale{0.5}
\plotone{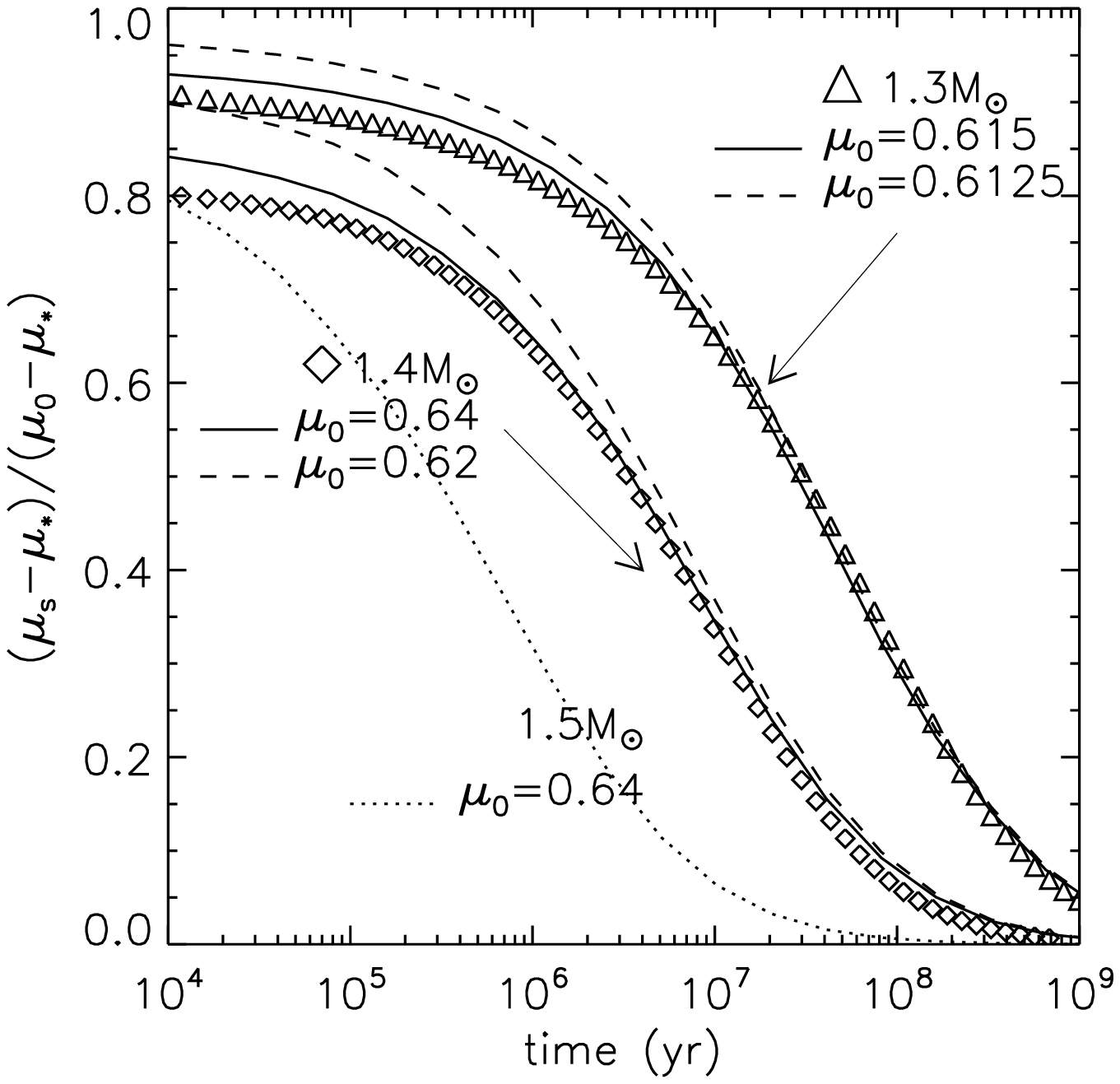}
\caption{Evolution of the relative enhancement in the surface mean molecular weight in the fingering phase. The symbols represents the numerical solutions, for the original 1.4$M_\odot$ example described in \S\ref{sec:example} and for the 1.3$M_\odot$ test case described in \S\ref{sec:estim}. The solid lines correspond to the estimated $\mu_s(t)$ for each case and fit the data very well. Dashed lines correspond to runs with lower initial enhancements $\mu_0$ as shown.  After about 10Myr, the evolution of the relative enhancement depends only on the stellar structure. Also shown as a dotted line is the {\it estimated} evolution of $\mu_s$ in a 1.5$M_\odot$ star. }
\label{fig:finger}
\end{figure}

The comparison of (\ref{eq:mus}) with the exact numerical solution is shown in Figure \ref{fig:finger}, for the 
original run with the 1.4$M_\odot$ star, and the additional test made with the 1.3$M_\odot$ star (see \S\ref{sec:estimdyn}). The value of $\alpha$ is adjusted to fit the numerical solution for the 1.4$M_\odot$ run: with $\alpha = 3.8$ the approximate solution fits the numerical solution very well. The same value of $\alpha$ can then be used to estimate $\mu_s(t)$ for the 1.3$M_\odot$ run, and comparison with the true solution is again excellent. In both cases the main source of discrepancy turns out to be the estimate of $\mu_1$, which can be off by about 10\%. 

Having validated the semi-analytical estimates, I now use them on a variety of initial conditions and stellar masses (see Figure \ref{fig:finger}). 
The main result is that, when viewed in terms of the {\it relative} enhancement $(\mu_s(t) - \mu_\star)/(\mu_0 - \mu_\star)$, the evolution of the surface mean molecular weight about 10Myr post-impact only depends on the structure of the host star and not on the initial conditions. The timescale for the relative enhancement to drop to 10\% strongly decreases with increasing stellar mass, from about 600Myr for a 1.3$M_\odot$ star, to 60Myr for a 1.4$M_\odot$ star and 6Myr for a 1.5$M_\odot$ star.  Hence, one would be exceedingly lucky to detect any excess metallicity through planetary pollution on a 1.4$M_\odot$ and 1.5$M_\odot$ star. Meanwhile, higher-than-average-metallicity on a star of mass 1.3$M_\odot$ or smaller could be the sign of a recent ($<1$Gyr old) event (e.g. HD 149026, see \citet{Lial2008}), but would of course require a very metal-rich, high-mass impactor to provide a detectably-high $\mu_0$. 

\section{Conclusion}
\label{sec:ccl}

Thanks to the remarkable scientific contribution of \citet{Traxleral2010}, we now have reliable universal laws for turbulent mixing by fingering convection. In this Letter I apply them to the question of whether the metallicity enhancement caused by the recent infall of a planet onto its host star can be detected. 

As suggested by \citet{vauclair2004}, an infall event first causes a dynamical readjustment of the outer regions, triggered as the adverse compositional gradient destabilizes the system following the Ledoux criterion. This partially dilutes the added material into the upper radiative zone. 
Then follows a longer mixing phase, in which the radiative zone becomes unstable to fingering convection. The induced turbulence dilutes the remainder of the metallicity excess into the radiative zone. The evolution of the surface metallicity in both phases can be modeled accurately through simple semi-analytical estimates provided in \S\ref{sec:estimdyn} and \S\ref{sec:estimfing}.

I find that the relative metallicity enhancement decreases on a timescale which, after the first few Myrs, {\it depends only on the stellar structure} and is much shorter for higher-mass stars ($\sim 10$Myr) than for lower-mass stars ($\sim 1$Gyr), see \S\ref{sec:estimfing} for detail. This timescale is much longer than the one estimated by \citet{vauclair2004}, but is nevertheless quite short. This mass-dependence compensates the initial condition trend, whereby the post-impact metallicity is inversely related to the convection zone mass for a given impactor. These results explain why no trend in metallicity vs. stellar mass has been detected in planet-bearing stars \citep{SIM2001,FischerValenti2005}. Furthermore, between the rapid dilution timescale for higher-mass stars and the  the larger convection zone mass of lower-mass stars, the statistically-higher metallicity observed in planet-bearing stars {\it cannot} be due to planetary pollution: it has to be a primordial effect. 

Finally, note that by contrast with the assumption of \citet{vauclair2004}, I find that the fingering region can extend very deeply into a star, and in particular, down to the Li-burning region. This would provide a simple theoretical backing to the observations of lower Li abundances in planet-bearing stars \citep{Gonzalez2008,Israelianetal2009}.




\begin{thebibliography}{18}
\expandafter\ifx\csname natexlab\endcsname\relax\def\natexlab#1{#1}\fi


\bibitem[{{Denissenkov}(2010)}]{Denissenkov2010}
{Denissenkov}, P.~A. 2010, ArXiv e-prints

\bibitem[{{Fischer} \& {Valenti}(2005)}]{FischerValenti2005}
{Fischer}, D.~A., \& {Valenti}, J. 2005, \apj, 622, 1102

\bibitem[{{Garaud} \& {Bodenheimer}(2010)}]{GaraudBodenheimer2010}
{Garaud}, P., \& {Bodenheimer}, P. 2010, \apj, 719, 313

\bibitem[{{Gonzalez}(2008)}]{Gonzalez2008}
{Gonzalez}, G. 2008, \mnras, 386, 928

\bibitem[{{Israelian} {et~al.}(2009){Israelian}, {Delgado Mena}, {Santos},
  {Sousa}, {Mayor}, {Udry}, {Dom{\'{\i}}nguez Cerde{\~n}a}, {Rebolo}, \&
  {Randich}}]{Israelianetal2009}
{Israelian}, G., {et~al.} 2009, \nat, 462, 189

\bibitem[{{Jackson} {et~al.}(2009){Jackson}, {Barnes}, \&
  {Greenberg}}]{JBG2009}
{Jackson}, B., {Barnes}, R., \& {Greenberg}, R. 2009, \apj, 698, 1357

\bibitem[{{Laughlin} \& {Adams}(1997)}]{LaughlinAdams1997}
{Laughlin}, G., \& {Adams}, F.~C. 1997, \apjl, 491, L51+

\bibitem[{{Levrard} {et~al.}(2009){Levrard}, {Winisdoerffer}, \&
  {Chabrier}}]{LWC2009}
{Levrard}, B., {Winisdoerffer}, C., \& {Chabrier}, G. 2009, \apjl, 692, L9

\bibitem[{{Li} {et~al.}(2008){Li}, {Lin}, \& {Liu}}]{Lial2008}
{Li}, S., {Lin}, D.~N.~C., \& {Liu}, X. 2008, \apj, 685, 1210

\bibitem[{{Lin} \& {Papaloizou}(1986)}]{LinPapaloizou1986}
{Lin}, D.~N.~C., \& {Papaloizou}, J. 1986, \apj, 309, 846

\bibitem[{{Santos} {et~al.}(2001){Santos}, {Israelian}, \& {Mayor}}]{SIM2001}
{Santos}, N.~C., {Israelian}, G., \& {Mayor}, M. 2001, \aap, 373, 1019

\bibitem[{Schmitt(1983)}]{schmitt1983}
Schmitt, R. 1983, Phys. Fluids, 26, 2373

\bibitem[{{Stern}(1960)}]{stern1960}
{Stern}, M.~E. 1960, Tellus, 12, 172

\bibitem[{{Torres} {et~al.}(2008){Torres}, {Winn}, \& {Holman}}]{Torresal2008}
{Torres}, G., {Winn}, J.~N., \& {Holman}, M.~J. 2008, \apj, 677, 1324

\bibitem[{{Traxler} {et~al.}(2010){Traxler}, {Garaud}, \&
  {Stellmach}}]{Traxleral2010}
{Traxler}, A.~L., {Garaud}, P., \& {Stellmach}, S. 2010, \apj

\bibitem[{{Ulrich}(1972)}]{ulrich1972}
{Ulrich}, R.~K. 1972, \apj, 172, 165

\bibitem[{{Vauclair}(2004)}]{vauclair2004}
{Vauclair}, S. 2004, \apj, 605, 874

\end{thebibliography}

\end{document}